\documentclass[10pt,twocolumn,showpacs,amsmath,pra,amssymb,a4paper]{revtex4}
\usepackage[T1]{fontenc}
\usepackage[latin9]{inputenc} 
\usepackage{graphicx}
\usepackage{amsbsy}
\usepackage{amssymb}
\usepackage[amssymb]{SIunits}
\usepackage{esint}
\usepackage{times}
\usepackage{color}
\usepackage{xspace}

\newcommand{\rsec}[1]{\emph{#1.} ---\xspace}
\newcommand{\unif}{translation\-ally-invariant\xspace}

\newcommand{\vl}[1][{\br}]{\ensuremath{V_\mathrm{latt}(#1)}\xspace}
\newcommand{\vtn}{\ensuremath{V_\mathrm{tr}}\xspace}
\newcommand{\vt}[1][{\br}]{\ensuremath{\vtn(#1)}\xspace}

\newcommand{\nt}{\ensuremath{\tilde{n}}\xspace}
\newcommand{\ntd}[2][{}]{\ensuremath{\tilde{n}_{#2}^{#1}\xspace}}
\newcommand{\Nt}{\ensuremath{\tilde{N}}\xspace}

\newcommand{\bose}[3][{}]{\ensuremath{\zeta^{#1}_{#2}\negthinspace\left(#3\right)}\xspace}
\newcommand{\bosee}[3][{}]{\ensuremath{\zeta^{#1}_{#2}\negthinspace\left(e^{#3}\right)}\xspace}
\newcommand{\boseze}[2]{\bose{#1}{z e^{#2}}}

\newcommand{\fracl}[2]{\ensuremath{{#1}/{#2}}\xspace} 
\newcommand{\fracb}[2]{\left(\frac{#1}{#2}\right)}
\newcommand{\fraclb}[2]{\left(\fracl{#1}{#2}\right)\xspace}  
\newcommand{\fraclbs}[2]{\left[\fracl{#1}{#2}\right]}

\newcommand{\kb}{\ensuremath{k}\xspace}
\newcommand{\kt}{\ensuremath{\kb T}\xspace}
\newcommand{\tcl}{\ensuremath{{T_L^0}}\xspace} 
\newcommand{\tcn}{\ensuremath{{T_c^0}}\xspace}
\newcommand{\ktcl}{\ensuremath{{\kb \tcl}}\xspace}
\newcommand{\ktc}{\ensuremath{{\kb T_c}}\xspace}
\newcommand{\tcfs}{\ensuremath{{T_c^{\mathrm{fs}}}}\xspace}
\newcommand{\ktcn}{\ensuremath{{\kb T_c^0}}\xspace}

\newcommand{\at}[2]{\left[#1\right]_{#2}\xspace}

\newcommand{\ob}{\ensuremath{\bar{\omega}}\xspace}
\newcommand{\hwb}{\ensuremath{\hbar\ob}\xspace}
\newcommand{\Kbm}[1][{b}]{\ensuremath{K_{#1}^{\mathrm{min}}}\xspace}
\newcommand{\half}[1][1]{\frac{#1}{2}\xspace}
\newcommand{\diff}{\mathrm{d}}
\newcommand{\dbr}{\diff\br}
\newcommand{\ns}{\ensuremath{N_s}\xspace} 
\newcommand{\tdim}{\ensuremath{d}\xspace}
\newcommand{\tdimtl}{\ensuremath{\fracl{\tdim}{2}}\xspace}
\newcommand{\zetaf}[2]{\zeta\hspace{-1mm}\left(\tfrac{#1}{#2}\right)\xspace}
\newcommand{\od}[1]{\ensuremath{{O\negthickspace\left({{#1}}\right)}}\xspace}
\newcommand{\bzero}{\ensuremath{\mathbf{0}}\xspace}
\newcommand{\mufs}{\epsilon_0}
\newcommand{\br}{\ensuremath{\mathbf{r}}\xspace}

\begin{document}
\title{Critical temperature of a Bose gas in an optical lattice}
\author{D. Baillie and P. B. Blakie}
\affiliation{Department of Physics, Jack Dodd Centre for Quantum Technology, University of Otago, P.O. Box 56, Dunedin, 9016 New Zealand}
\pacs{67.85.Hj, 03.75.Hh, 05.30.Jp }

\date{\today}
\begin{abstract}
We present theory for the critical temperature of a Bose gas in a combined harmonic lattice potential based on a mean-field description of the system. 
We develop practical expressions for the ideal-gas critical temperature, and corrections due to interactions, the finite-size effect, and the occupation of excited bands.  We compare our expressions to numerical calculations and find excellent agreement over a wide parameter regime. 
\end{abstract}

\maketitle 
Ultra-cold atoms in optical lattices have emerged as a flexible system for studying many-body physics \cite{Jaksch98,Greiner02a,Lewenstein07}. Despite immense interest in this system, our understanding of the fundamental phenomenon of Bose-Einstein condensation, in particular the critical temperature, $T_c$, is rudimentary compared to the case of the harmonically trapped gas. Much of the difficulty in making predictions for the lattice system arises from the complex spectrum \cite{Hooley04,Rey06,Blakie07a} in the combined harmonic lattice (CHL) potential used in experiments. In contrast, the power-law density of states for systems in purely harmonic traps has led to simple expressions for $T_c$ \cite{Bagnato87a}, and the finite-size \cite{Grossmann95} and (mean-field) interaction  \cite{Giorgini96} corrections. 
Although in combination these corrections are at the $\sim$10\% level for the harmonically trapped gas, beautiful experiments by the Orsay group \cite{Gerbier04} have been able to exclude ideal-gas behavior by more than 
two standard deviations, and find quantitative agreement with mean-field theory. The prospects for similar thermodynamic studies in optical lattice experiments has taken a leap forward with the recent demonstration of a practical scheme for measuring temperature in this system  \cite{McKay09} (see also \cite{Catani09a}). 

Here, we first derive an expression for the ideal-gas critical temperature, $T_c^0$, of the experimentally relevant CHL system by introducing a \textit{shape approximation} to the lattice band structure. By explicitly including the physics of the finite width of the ground band, our result demonstrates remarkably good agreement with the exact (ideal) result \cite{Blakie07a}.  
 We then derive expressions for the critical temperature corrections due to finite size, excited band, and mean-field interaction effects. 
 We compare our expressions to full self-consistent three-dimensional (3D) numerical calculations using the actual band structure and demonstrate their excellent agreement over a broad range of experimentally relevant parameters. We observe that the interaction correction is most significant, heavily suppressing $T_c$. We find that the validity range of the quadratic shape approximation is complementary to the effective-mass approximation, so that simple descriptions extend over a wide range. All of our expressions depend only on experimental parameters and quantities that can be simply evaluated from the single-particle spectrum of the translationally-invariant lattice (TIL).

The general CHL potential we consider is $\vl+\vt$, for the case of $1\le d\le3$\label{p:d} spatial dimensions. The lattice potential is given by $\vl \equiv \sum_{j=1}^d V_j\sin^2\left(\fracl{\pi r_j}{a_j}\right)$, where $V_j$ is the lattice depth and $a_j$ is the lattice site spacing, in direction $j$.  The harmonic trap is $\vt \equiv \half m \sum_{j=1}^d  \omega_j^2 r_j^2$\label{p:vtr}, where $m$ is the atomic mass and $\omega_j$ is the harmonic trap frequency. 
  We also consider the TIL case, where $\vt=0$. We note that while all the numerical results we present are for the cubic lattice case (all  $a_j=a$, $V_j=V$), our expressions also apply to non-cubic lattices.
 
The Hartree-Fock description of bosons in an optical lattice at $T\ge T_c$ gives the per-site density of atoms in band $b$ \cite{fillingref}
\begin{align}
  \nt_b(\br) &= \int\frac{a^d g_b(K)\,\diff K }{e^{\beta(K + \vt + 2{ \sum_{b'} U_{bb'} \nt_{b'}(\br)} - \mu)}-1}.\label{e:nbr}
\end{align}
We have used the local density approximation, valid for $\kt \gg \hbar \ob^*$, where $\ob^* \equiv \sum_j \omega^*_j/d$, $\omega_j^* \equiv \omega_j\sqrt{m/m^*_j}$ and $m^*_j$ is the ground-band effective mass along direction $j$ (evaluated at $\mathbf{0}$ quasi-momentum). Here, $g_b(K)$ is the density of states of the corresponding TIL (i.e.~usual Bloch spectrum) and the interaction coefficient  $U_{bb'}$ is proportional to the $s$-wave scattering length  $a_s$ and the integral of the product of Wannier state densities for bands $b$ and $b'$   \cite{Baillie09b} (also see \cite{Lin08}). We have also introduced $\beta=1/kT$ and the chemical potential, $\mu$. 
Equation \eqref{e:nbr}, derived from an extended Bose-Hubbard Hamiltonian, becomes ambiguous in the shallow lattice limit ($V\lesssim2E_R$) where the Wannier states become delocalized. We implement a continuation of the theory into this regime (see also \cite{Baillie09b}), but note that very shallow lattices are best examined without making a band decomposition \cite{Zobay06}. 
We also note that mean-field theory is invalid for interacting systems in the deep lattice limit where correlations are important \cite{Jaksch98,Greiner02a}.   The breakdown of mean-field theory  has been studied for a one dimensional optical lattice at $T=0$ \cite{Rey03}. For experiments with $^{87}$Rb, \eqref{e:nbr} is a good description for $V\lesssim13E_R$. 

The total number of thermal atoms in band $b$ is $\Nt_b = a^{-d}\int\dbr \,\nt_b(\br)$. 
For a wide parameter regime of interest, the critical point occurs when the occupation of excited bands is negligible. Initially, we consider the ground band ($b=0$) in isolation and assume it contains all atoms in the system (i.e.~$\Nt_0=N$).
We will later account for the influence of excited bands. 

\begin{figure}[t]
\includegraphics{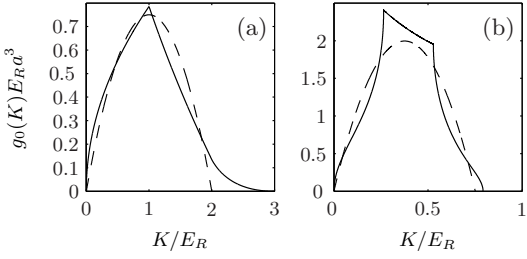}
  \vspace{-0.4cm}\caption{Exact (solid) and quadratic shape approximation (dash) 3D cubic TIL ground-band density of states for (a) $V=0$ and (b) $V=5E_R$, where $V\equiv \prod_j V_j^{1/d}$, $E_R \equiv h^2/8ma^2$ and $a\equiv \prod_j a_j^{1/d}$.}
  \label{f:quaddos}\vspace{-0.5cm}
\end{figure}
\begin{figure}[t]
   \hspace{-0.58cm}\includegraphics{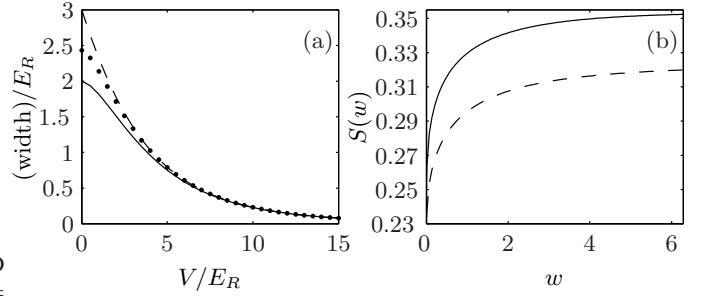}
  \vspace{-0.5cm}\caption{
  (a) Width of the 3D cubic TIL ground band, using \eqref{e:Wdef} (solid), maximum less minimum energies (dash) and the tight-binding approximation \cite{Jaksch98,Blakie04} (dot).  
  (b) Spread function $S(w)$ using \eqref{e:swcendelta} (solid) and the quadratic shape approximation with \eqref{e:swdef} (dash). 
\vspace{-0.4cm}}
  \label{f:SandWidth}
\end{figure}

The bandwidth is an important parameter, which for band $b$, we calculate as
\begin{align}
  \frac{W_b}{2} \equiv \left[a^d\int K g_b(K)\diff K\right] - \Kbm \label{e:Wdef},
\end{align}
where $\Kbm$ is the minimum energy of the band. This definition for $W_b$ provides a useful characterization of the bandwidth in shallow lattices (e.g.~see Fig.~\ref{f:quaddos}), 
approaches the tight-binding expression in deep lattices (see Fig.~\ref{f:SandWidth}(a)), and extends our treatment to non-cubic lattices.

\rsec{Ideal-gas critical temperature} For sufficiently deep lattices, the TIL ground bandwidth ($W\equiv W_0$) becomes negligible compared to other relevant energy scales. Then, the exact structure of the ground band becomes unimportant and band shape approximations can be made.    
As a first approximation, we set $g_0(K)=\delta(K-W/2)/a^d$. Neglecting interactions, \eqref{e:nbr} gives $\ntd{0}(\br) = \left\{z^{-1}\exp[\beta\vt+w/2] - 1\right\}^{-1}$, where we have defined the \textit{thermal bandwidth}  $w\equiv W/\kt$, and $z\equiv e^{\beta\mu}$. Then $\Nt_0 = \ns \boseze{\tdimtl}{-w/2}$  where 
$\bose{\alpha}{z} \equiv \sum_{n=1}^\infty \fracl{z^n}{n^\alpha}$ is the Bose function,
$d$ is the number of harmonically trapped dimensions (i.e. $d=0$ for the TIL), 
$\ns$ is the number of sites for the TIL and 
$\ns=(2\pi \kt /m\omega^2a^2)^{d/2}$ is a measure of the thermally accessible number of sites for the CHL, with $\omega\equiv \prod_j \omega_j^{1/d}$. In the limit $w\rightarrow 0$ \cite{cdeltaref}, for $d<3$, or for the TIL, there is then no condensation, and for the 3D CHL, the critical temperature is 
\begin{align}
  \tcl =  \frac{m\omega^2 a^2}{2\pi\kb}\left[\frac{N}{\zetaf{3}{2}}\right]^{2/3},\label{e:tclocal}
\end{align}
with $\zeta(\alpha) \equiv \bose{\alpha}{1}$, as in \cite{Blakie07a} (also see \cite{Muradyan08}). 
A better approximation to $g_0(K)$ is to use the quadratic shape approximation, $g_0(K) = 6K(W-K)/W^3a^d$ for $0<K<W$ and zero otherwise \cite{rectriref}, as shown in Fig. \ref{f:quaddos}. Using \eqref{e:nbr}, this gives\enlargethispage{1cm}
\begin{align}
\nt_0(\br) 
      &= \frac{6}{w^3}\left[ w \boseze{2}{-\beta\vt} - 2 \boseze{3}{-\beta\vt} \right.\notag\\&\left.+ w \boseze{2}{-\beta\vt-w} + 2 \boseze{3}{-\beta\vt-w}\right],\label{e:ntquad}\\
\Nt_0   &= \frac{6\ns}{w^3}\left[ w \bose{2+\tdimtl}{z} - 2 \bose{3+\tdimtl}{z} \right.\notag\\&\hspace{1.0cm}\left.+ w \boseze{2+\tdimtl}{-w} + 2 \boseze{3+\tdimtl}{-w}\right].\label{e:Ntquad}
\end{align}
For the 3D TIL this result gives $\tcn \approx W(N/\ns+\half)/3\kb$ \cite{tilref}. 
For the 3D CHL, the series expansion of \eqref{e:Ntquad}, convergent for $w<2\pi$ is \cite{Robinson51}
\begin{align}
 \Nt_0 &= \ns\left[\zetaf{3}{2} -\frac{48\sqrt{\pi w}}{35}   - \frac{\zetaf{1}{2}w}{2} + \od{w^{2}}\right].
\end{align}
Solving this implicit Taylor series, we find the ideal-gas critical temperature (with  $w^0_L \equiv \fracl{W}{\ktcl}$; the notation for the thermal bandwidth, uses the subscripts and superscripts from the corresponding temperature) 
\begin{align}
  \frac{\tcn-\tcl}{\tcl} &= \frac{32\sqrt{\pi}}{35\zetaf{3}{2}}\sqrt{w^0_L} 
   + \frac{\zetaf{3}{2}\zetaf{1}{2} + \frac{2304\pi}{1225}}{3\zetaf{3}{2}^2} w^0_L \notag\\
   &\hspace{-2.8mm}+ 32\sqrt{\pi}\frac{\zetaf{3}{2}\zetaf{1}{2} + \frac{256\pi}{245}}{105\zetaf{3}{2}^3} \left(w^0_L\right)^{3/2} + \od{w^0_L}^2\label{e:quadtcnall}\\
  &\approx 0.620
  \sqrt{w_L^0}
   + 0.102
    w_L^0 
   - 0.016 \left(w_L^0\right)^{3/2}.\label{e:quadtcn}
\end{align}

\rsec{Interaction shift}With harmonic confinement, the leading order effect of interactions on $T_c$ is at the mean-field level in 3D:  repulsive interactions reduce the density at trap center and decrease $T_c$. 
Our approach, similar to  \cite{Giorgini96}, involves a Taylor series expansion about the critical point. However, in the lattice, the ideal gas has divergent density at $\br=\mathbf{0}$ as $w\rightarrow 0$, from \eqref{e:ntquad} and \cite{Robinson51}. 
The effect of a non-zero interaction parameter, $U=U_{00}$, is significant at removing this near-divergent behavior. 
Thus, making the Taylor expansion at \tcn would significantly overstate the interaction effect, particularly for deep lattices. We therefore use the first terms in a Taylor series about the interacting critical temperature, $T_c$, (the quantity we aim to determine) 
and obtain, to first order in $\mu$ and $U$
\begin{align}
 \delta T_c \equiv T_c - \tcn &\approx 
 -\fracl{( \mu\partial_\mu\Nt_0 + U\partial_U \Nt_0) }{\partial_T{\Nt_0}},
\end{align}
where  $U \partial_U{\Nt_0} \approx -2U\int\dbr\,\nt_0(\br)\partial_\mu\nt_0(\br)/a^3$ and $\mu\rightarrow 2U\nt_0(\bzero)$, with the derivatives ($\partial_{\lambda}\!\equiv\!\frac{\partial}{\partial \lambda}$) and $\nt_0(\bzero)$ evaluated at $\mu=0,U=0$, and $T=T_c$. We obtain
\begin{align}
 \delta T_c &\approx -2U[1-S(w_c)] \nt_0(\bzero)\fracl{\partial_\mu \Nt_0}{\partial_T{\Nt_0}},\label{e:dtcf}
\end{align}
where $w_c \equiv W/\ktc$ and we have defined the \textit{spread function}
\begin{equation}
  S(w) \equiv \at{\frac{ \int\dbr\,\nt_0(\br) \partial_\mu{\nt_0(\br)} }{ \int\dbr\, \nt_0(\bzero) \partial_\mu{\nt_0(\br)}  }}{\mu=0}.\label{e:swdef}
\end{equation}
Using the approximation $g_0(K)=\delta(K-W/2)/a^3$, we find
\begin{align}
   S(w) &= \frac{e^{w/2}-1}{2}\left[ \frac{\bosee{-\half}{-w/2}}{\bosee{\half}{-w/2}}-1\right].\label{e:swcendelta}
\end{align}
For the quadratic shape approximation,  as $w\rightarrow\infty$, $S(w)\rightarrow \sum_{j,k=1}^\infty j^{-2} k^{-1} (j+k)^{-3/2}/{\zeta(2)\zetaf{5}{2}} \approx 0.325$, and we have evaluated $S(w)$ numerically as shown in Fig.~\ref{f:SandWidth}(b), which shows that $0.24 < S(w) < 0.325$
and, as $w\rightarrow 0$, $S(w)\rightarrow 0.24$. 
We note that for the (no lattice) 3D harmonic trap case,  
$S=0.281$ \cite{Giorgini96}.

We evaluate the derivatives required by \eqref{e:dtcf} analytically, then expand using \cite{Robinson51} to find (to first order in $U$)
\begin{align}
   \delta T_c &\approx 
   -\frac{32\sqrt{\pi}U[1-S(w_c)]}{5\zetaf{3}{2}\kb w_c^{3/2}} \approx - 3\frac{U}{\kb}\fracb{\ktc}{W}^{3/2},\label{e:intshiftapprox}
\end{align}
where the last result uses the approximation $S(w_c) \approx 0.3$, giving a simple equation to be solved for $T_c$.

\begin{figure}
  \hspace{-0.5cm}
  \includegraphics{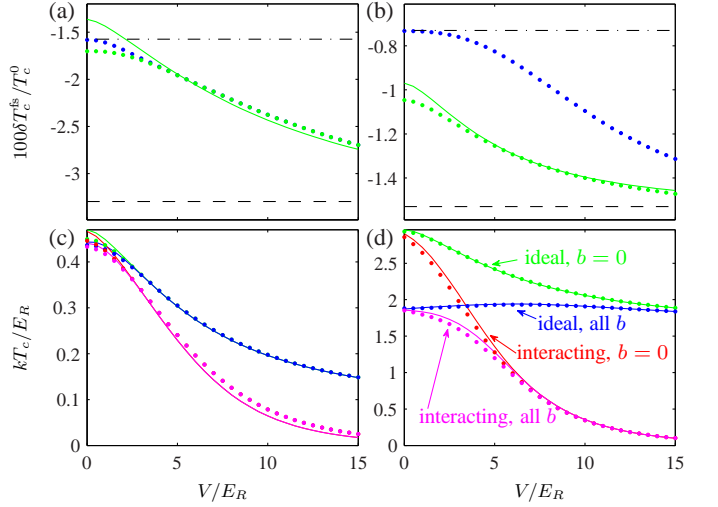}
    \vspace{-0.6cm}\caption{(Color) 3D cubic CHL results. (a,b) Comparison of ideal finite-size effect between quadratic shape approximation (solid) and numerical results (ground band in green and all bands in blue), effective-mass (\ref{e:fsemass}) (dash-dot) and $w\rightarrow 0$ limit (\ref{e:quadfsa}) (dash). (c,d) Comparison of quadratic shape approximation (solid) to numerical results (dot), using $a=425\:\nano\metre$, $a_s=5.77\:\nano\metre$ and ignoring the finite-size effect. Both ideal (ground band in green, all bands in blue) and interacting results (ground band in red, all bands in magenta) are shown. For these parameters, the Mott insulator transition for unit occupancy is at $V\approx 13E_R$.
(a,c) $\omega=2\pi\times32\,\hertz$, $N=10^5$. (b,d) $\omega=2\pi\times64\,\hertz$, $N=10^6$.}\vspace{-0.6cm}
  \label{f:allres}
\end{figure}

\rsec{Finite-size effect}
Accounting for the non-zero energy of the ideal ground state (and hence the saturated ideal-gas chemical potential) due to confinement, $\mufs= 3\hbar\ob^*/2$, gives rise to the so called finite-size shift in the condensation temperature \cite{Grossmann95}.
To quantify this effect for the 3D CHL we obtain 
$\delta \tcfs \approx -\fracl{\mufs\partial_\mu\Nt_0}{\partial_T{\Nt_0}}$ yielding
\begin{align}
   \delta \tcfs &\approx -\frac{\mufs}{\kb}\left[
   \frac{16\sqrt{\pi}}{15\zetaf{3}{2}\sqrt{w_c^0}}
   + \frac{2\zetaf{1}{2}}{3\zetaf{3}{2}} + \frac{512\pi}{525\zetaf{3}{2}^2}\right]\notag\\
  &\approx -\frac{\mufs}{\kb}\left[\frac{0.72}{\sqrt{w_c^0}} + 0.076\right],\label{e:quadfs}
\end{align}
where $w_c^0 \equiv W/\ktcn$.
	For a cubic lattice, using the tight-binding effective mass, we have $\ob^*=\ob\pi\sqrt{W/12E_R}$ \cite{Rey05} and $\mufs = \sqrt{3W/E_R}\pi\hwb/4$ where $\ob = \sum_j \omega_j/d$. In this limit, the highest order finite-size shift in \eqref{e:quadfs} depends on $W$ only through \tcn
\begin{align}
     \delta \tcfs &\approx -\frac{4\pi^{3/2}\hwb}{5\sqrt{3}\zetaf{3}{2}\kb}\sqrt{\frac{\ktcn}{E_R}}
.\label{e:tbfsa}
\end{align}
In the $W\rightarrow 0$ limit, we have, using \eqref{e:tclocal}
\begin{align}
     \frac{\delta \tcfs}{\tcl}  
     &\approx -  \frac{8\pi}{5\sqrt{3}\zetaf{3}{2}^{2/3}} \frac{\ob}{\omega} N^{-1/3}
     \approx -  1.5 \frac{\ob}{\omega} N^{-1/3}.\label{e:quadfsa}
\end{align}
We compare the finite-size effect using the quadratic shape approximation to the full numerical calculation (from \eqref{e:nbr} with $\mu=\mufs$, limiting the integral to $K+\vt > \epsilon_0$) in Fig.~\ref{f:allres}(a,b). 

\rsec{Excited bands}
Over a wide regime of experimental interest, the excited bands are weakly occupied at the critical point (i.e.~$kT_c < \Kbm$ for $b>0$). 
It is then sufficient to account for the small occupation using a simple band shape for excited bands, $g_b(K) = 1/W_ba^d$ for $0<K-\Kbm<W_b$ and zero otherwise, and neglecting excited-band interactions. We find the total excited band occupation (which acts to reduce the ground-band occupation from $N$) at temperature $T$
\begin{align}
\Nt_{\rm{ex}}(T)= \ns\sum_{b>0} \frac{kT}{W_b} e^{-\Kbm/kT}\left(1- e^{-W_b/kT} \right),\label{e:exbandadj}
\end{align}
where we have used the approximation $\bosee{\alpha}{-x} \approx e^{-x}$ for $x \gg 1$.  If $\Nt_{\rm{ex}}(T_c)\ll N$   (noting $T_c$ is often much smaller than $\tcn$  \eqref{e:intshiftapprox}), we may safely neglect excited bands. If excited band contributions are important, it is necessary to use a consistent procedure to determine $T_c$. The approach we use is:\\ 
{\bf 1}.~Calculate an initial estimate  $\tcn$ using \eqref{e:quadtcn} if $w_c^0\ll 2\pi$ (otherwise, solve \eqref{e:Ntquad}).\\
{\bf 2}.~Calculate the ground-band interaction shift, $\delta T_c$, using \eqref{e:intshiftapprox}. {\bf 3}.~Adjust the ground-band number $\Nt_0=N-\Nt_{\rm{ex}}(\tcn + \delta T_c)$.\\
{\bf 4}.~Recalculate   $\tcn$ using $\Nt_0$. \\
Finally, steps 3 and 4 are iterated to find $\tcn$ if necessary. 
The critical temperature allowing for interactions in the ground band, with an adjustment for  excited bands is then $\tcn + \delta T_c$.

\rsec{Results}
We have verified the usefulness of our expressions by comparison with full self-consistent numerical calculations using the actual band structure, (\ref{e:nbr}), over a wide range of experimentally relevant parameter regimes.
The ideal-gas results in Fig.~\ref{f:allres}(a,b) reveal that the finite-size correction is typically only a few percent of $\tcn$, and is usefully bounded by the limiting expressions   \eqref{e:quadfsa} and \eqref{e:fsemass}. Fig.~\ref{f:allres}(c,d) examines the more significant interaction and excited band effects. The system in Fig.~\ref{f:allres}(c) is typical of current experiments and reveals the large suppression of $T_c$ from $\tcn$ due to interactions, and negligible contribution from excited bands. The system considered in Fig.~\ref{f:allres}(d), with larger atom number and tighter harmonic confinement, has higher critical temperature giving appreciable excited band effects \cite{fourbandsref}, especially in the shallow lattice regime ($V\lesssim4E_R$). 
We observe, from the full range of results in Fig.~\ref{f:allres}, that our expressions are in good agreement with the numerical calculations.

\begin{figure} 
  \includegraphics{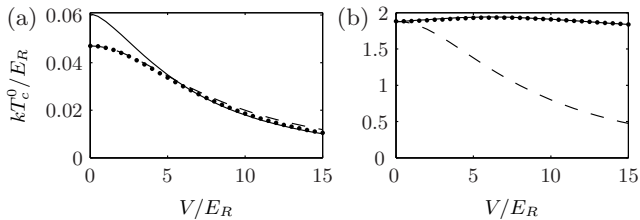}
  \vspace{-0.4cm}\caption{Comparison of actual (dot) to effective-mass (dash) and quadratic shape approximation (solid) critical temperature for an ideal Bose gas in a 3D cubic CHL with
   (a) $\omega=2\pi\times 16\,\hertz$, $N=10^3$, 
   (b) $\omega=2\pi\times 64\,\hertz$, $N=10^6$.
}\vspace{-0.6cm}
  \label{f:emassres}
\end{figure}

\rsec{Effective mass}
In the low temperature regime ($\ktc \ll W$) the quadratic shape approximation is less effective  as the detailed features of $g_0(K)$ at low $K$ become important. 
 However, for an ideal gas in this regime we can employ an effective mass approximation. 
For the TIL case, this predicts
no condensation for $d=1,2$, and yields 
     $\tcn = \fraclb{2\pi\hbar^2}{\kb m^*} \fraclbs{N}{\ns a^3 \zeta(3/2)}^{2/3}$  in 3D   \cite{Kleinert04}. 
For the CHL, there is no condensation for $d=1$ and for $d=2,3$, we have 
  $\tcn = \hbar\omega^*\fraclbs{N}{\zeta(d)}^{1/d}/\kb$ \cite{Bagnato87a}. 
We compare the effective-mass predictions to our full numerical calculation for a 3D CHL in Fig.~\ref{f:emassres}(a), showing good agreement in the low temperature regime, although the effective-mass approximation is usually very poor except for very shallow lattices (e.g. Fig.~\ref{f:emassres}(b)). 
The finite-size effect for the 3D CHL, using the methodology of \cite{Grossmann95}, 
is
\begin{align}
\frac{\delta \tcfs}{\tcn} \approx  -\frac{\ob^*}{\omega^*}\frac{\zeta(2)}{2\zeta(3)^{2/3}} N^{-1/3} \approx -0.73\frac{\ob^*}{\omega^*}N^{-1/3}, \label{e:fsemass}
\end{align}
as shown on Fig.~\ref{f:allres}(a,b) (for a cubic lattice ${\ob^*}/{\omega^*}={\ob}/{\omega}$).

\rsec{Conclusions}
We derived practical expressions for the ideal and mean-field interacting critical temperature in an optical lattice by using a shape approximation for the density of states. We have given primary consideration to the experimentally relevant case of the 3D combined harmonic lattice potential, but have also given results for the \unif lattice. For the ideal case we have shown that our validity range (low to moderate values of $w$) is complementary to that of the effective-mass approximation (large values of $w$). By using the average energy of the band to calculate the width in \eqref{e:Wdef}, we have extended the validity range into shallow lattices.

We have compared all of our results to full self-consistent numerical calculations with the actual band structure and have generally found excellent agreement over a wide parameter regime.
We have derived a formula for the finite-size effect in the lattice, and the important effect of mean-field interactions in the ground band. We have also demonstrated a  procedure to include the influence of excited bands.

Our results indicate that, for the typical parameters of current optical lattice experiments,  interaction shifts in $T_c$ are large compared to the harmonically trapped case (where the interaction shift is usually only a small fraction of $\tcn$ \cite{Giorgini96,Gerbier04}), and that excited band effects are small. Our main results, contained in \eqref{e:quadtcnall}, \eqref{e:intshiftapprox} and \eqref{e:quadfs}, are easy to evaluate,  depending only on experimental parameters and the width of the ground band. Our theory is applicable to non-cubic lattices with appropriate identification of the bandwidth parameter \eqref{e:Wdef}. With the recent progress in measuring temperature in lattices we believe our results will be directly comparable to experiments in the near future. 

The authors acknowledge support from  the University of Otago Research Committee and NZ-FRST contract NERF-UOOX0703, and useful discussions with Ashton Bradley.


\begin{thebibliography}{20}
\bibitem{fillingref}\bibinfo{note}{As defined,  $\tilde{n}_b(\br)$ is the filling factor envelope function, with $\tilde{n}_b(\br)a^{-d}$ the coarse-grained density \cite{Baillie09b}.}
\bibitem{cdeltaref}\bibinfo{note}{The form $g_0(K)=\delta(K)/a^d$ differs from $g_0(K)=\delta(K-W/2)/a^d$, because the energy origin is set by $\mu\rightarrow 0$.}
\bibitem{rectriref}\bibinfo{note}{Rectangular and triangular band shapes can also be used but the quadratic shape gives better results.}
\bibitem{tilref}\bibinfo{note}{The TIL is sensitive to the details of $g_0(K)$ and for $d=1,2$ the quadratic shape approximation is poor due to Van Hove singularities at $K=0$.}
\bibitem{fourbandsref}\bibinfo{note}{Fifth and higher bands are treated as non-interacting.}
\end{thebibliography}
\end{document}